\documentclass[12p,preprint]{revtex4}
\usepackage[english]{babel}
\usepackage{pgf,pgfarrows,pgfnodes,pgfautomata,pgfheaps,subfigure}
\usepackage{amsmath,amssymb}
\usepackage[latin1]{inputenc}

\begin{document}
\title{On the  partial trace over collective spin-degrees of freedom}
\author{Yamen Hamdouni}

\email{hamdouniyamen@gmail.com}

\affiliation {School of Physics, University of  KwaZulu-Natal,
Westville Campus, Durban 4001, South Africa}

\begin{abstract}
We derive analytical properties for the degeneracy $\nu(N,j)$
occurring in the decomposition
$\bigoplus\limits_{j}^\frac{N}{2}\nu(N,j)\mathbb
 C^{2j+1}$ of the state space $\mathbb C^{2\otimes N}$. We also
investigate the dynamics of two qubits coupled  via Ising
interactions to separate spin baths, and we study  the thermodynamic
limit.
\end{abstract}

\maketitle

%\begin{pacs} 03.65.Yz, 03.67.Lx, 73.21.La, 75.10.Jm.}

 In recent years
there has been an increasing interest in the description of the
dynamics of small quantum systems interacting with their
surrounding~\cite{1}. This was motivated by the necessity of
 understanding  the phenomenon of decoherence in  quantum
systems~\cite{2,3,4,5}, and the attempt to build quantum devices
that enable the implementation of  quantum algorithms~\cite{6}.
However, the main difficulty one faces in such a task consists in
dealing with the large number of environmental degrees of freedom,
which makes most of  the proposed theoretical models impossible to
be solved analytically even for finite sizes of the surrounding.

Among the  promising candidates to quantum information processing
and quantum computing, spin systems seem to be the most suitable for
the construction of quantum gates~\cite{7,8}. Recently, it has been
shown that exact analytical solutions can be obtained for the
dynamics of few central qubits coupled to spin  baths of finite and
infinite sizes~\cite{9,10,11}. There, the interaction Hamiltonians
together with the baths Hamiltonians are functions of the collective
spin operators of the environments. In order to derive the reduced
density matrix of the central qubits, the partial trace over the
environmental spin degrees of freedom was carried out within the
subspaces corresponding to the different  values of the total
angular momentum of the surrounding.

Recall that the state space of single spin-$\frac{1}{2}$ particle is
 given by ${\mathbb C}^2$, where $\mathbb C$ denotes the field
of complex numbers. The corresponding basis is formed by the
eigenvectors $ \{|-\rangle,|+\rangle\}$ associated with the
eigenvalues $\pm\frac{1}{2}$ of the operator
$S_z=\frac{1}{2}\sigma_z$, where $\sigma_z$ designates the
$z$-component of the Pauli operator $\vec{\sigma}$.  In  general,
the state space of a system of $N$ qubits is given by the $N$-fold
tensor product of the state spaces of the individual particles,
namely, $\mathbb C^{2\otimes N}$. One possible basis of the latter
space consists of the  state vectors
$\bigotimes_i^N|\epsilon_i\rangle$, with $\epsilon_i=\pm$. These are
eigenvectors of the collective spin operator $J_z$, where
$\vec{J}=\frac{1}{2}\sum_{i=1}^N\vec{\sigma_i}$. Alternatively, one
can construct new basis composed of the common eigenvectors of the
operators $J^2$ and $J_z$; we shall denote them by $|j,m\rangle$
such that $\kappa\le j\le N/2$ and $-j\le m \le j$, as imposed by
the laws of addition  of angular momentum in quantum
mechanics~\cite{12}. In the above, $\kappa=0$ for $N$  even, and
$\kappa=1/2$ for $N$ odd. Note that the scalar product of state
vectors corresponding to different values of $j$ vanishes. This
means that the total  space $\mathbb C^{2\otimes N}$ can be
decomposed as the direct sum of subspaces ${\mathbb C}^{2j+1}$, that
is
\begin{equation}
 \mathbb C^{2\otimes N}=\bigoplus\limits_{j=\kappa}^\frac{N}{2}\nu(N,j)\mathbb
 C^{2j+1}.\end{equation}
The quantity $\nu(N,j)$ is the multiplicity corresponding to the
value $j$ of the total angular momentum; its exact form
reads~\cite{13}

\begin{equation}
\nu(N,j)=\binom{N}{N/2-j}-\binom{N}{N/2-j-1}=\frac{2j+1}{\frac{N}{2}+j+1}
\frac{N!}{(\frac{N}{2}-j)! (\frac{N}{2}+j)!}.
\end{equation}
Hence, given any operator $\hat{G}(\vec{J})$ on $\mathbb C^{2\otimes
N}$, its trace can be written as
\begin{equation}
\mathrm{tr} \ \hat{G}=
\sum\limits_{j=\kappa}^{\frac{N}{2}}\nu(N,j)\sum\limits_{m=-j}^j\langle
j,m|\hat{G}|j,m\rangle.\label{tr1}\end{equation}

Following the general ideas of the theory of open quantum systems,
the problem of finding a relation between the multiplicities of the
subspaces $\mathbb C^{2\otimes N_i}$ and that of $\mathbb
C^{2\otimes N}$, where $\sum_i N_i=N$, naturally arises. In this
work we illustrate how this problem can be solved, in the case
$N=N_1+N_2$, using  the invariance of the trace. The latter property
will also be used to describe the dynamics of two qubits in separate
spin baths.

 \subparagraph*{A decomposition law for the degeneracy
$\nu(N,J)$.} Let us  denote by $|j_i,m_i\rangle$ the basis state
vectors in the space $\mathbb C^{2\otimes N_i}$ ($i=1,2$). Hence the
trace of $\hat{G}(\vec{J})$ can also be  expressed as
\begin{equation}
\mathrm{tr} \hat
G=\sum_{j_1=\kappa_1}^{N_1/2}\sum_{m_1=-j_1}^{j_1}\sum
_{j_2=\kappa_2}^{N_2/2}\sum_{m_2=-j_2}^{j_2}\nu(N_1,j_1)\nu(N_2,j_2)\langle
j_1,j_2,m_1,m_2|\hat G|j_1,j_2,m_1,m_2\rangle.\label{tr2}
\end{equation}
On the other hand we have~\cite{14}
\begin{align}
|j_1,j_2,m_1,m_2\rangle=\sum_{J=|j_1-j_2|}^{j_1+j_2}\sum_{M=-J}^{J}
(-1)^{j_1-j_2+M}\sqrt{2J+1}\nonumber \\
\times
\begin{pmatrix}j_1&&j_2&&J\\m_1&&m_2&&-M\end{pmatrix}|J,M\rangle,\label{tr3}
\end{align}
where the quantity in matrix form denotes Wigner $3j$-symbol;
obviously,  the condition $m_1+m_2=M$ along with the triangle rule
$|j_1-j_2|\le J \le j_1+j_2$ must be satisfied. By
equations~(\ref{tr2}) and (\ref{tr3}), we can write:
\begin{align}
\mathrm{tr} \hat G=\sum_{j_1,m_1}\sum_{j_2,
m_2}\nu(N_1,j_1)\nu(N_2,j_2)\sum_{J,J'=|j_1-j_2|}^{j_1+j_2}
\sum_{M=-J}^J\sum_{M'=-J'}^{J'}(-1)^{2(j_1-j_2)+M+M'}\nonumber\\
\sqrt{(2J+1)(2J'+1)}
\begin{pmatrix}j_1&&j_2&&J\\m_1&&m_2&&-M\end{pmatrix}\begin{pmatrix}
j_1&&j_2&&J'\\m_1&&m_2&&-M'\end{pmatrix}\langle J',M'|\hat
G|J,M\rangle,\label{trn1}
\end{align}
where we have used the fact that $3j$-symbols are real. The operator
$\hat G$ is arbitrary; it can be chosen such that it satisfies
$\langle J',M'|\hat G|J,M\rangle=\langle J,M|\hat G|J,M\rangle
\delta_{JJ'}\delta_{M M'}$. In this case equation~(\ref{trn1})
reduces to
\begin{align}
\mathrm{tr} \hat G=\sum_{j_1,m_1}\sum_{j_2,
m_2}\nu(N_1,j_1)\nu(N_2,j_2)\sum_{J=|j_1-j_2|}^{j_1+j_2}
\sum_{M=-J}^J (-1)^{2(j_1-j_2)+2M}\nonumber\\
(2J+1)\Bigl\{\begin{pmatrix}j_1&&j_2&&J\\m_1&&m_2&&-M\end{pmatrix}\Bigl\}^2\langle
J,M|\hat G|J,M\rangle.\label{trn2}
\end{align}
The lower and upper limits of the sum over $J$ in the above equation
are, respectively, $|j_1-j_2|$ and $j_1+j_2$. For $J< |j_1-j_2|$, or
$J>j_1+j_2$, the  triple $(j_1,j_2,J)$ does not satisfy the triangle
rule and hence the  corresponding Wigner $3j$-symbol vanishes.
Consequently,  the right-hand side of equation~(\ref{trn2}) will not
be affected if we take $\frac{N_1+N_2}{2}$ as an upper limit, and
$\kappa$ as a lower limit for the sum over $J$   such that
$\kappa=0$ for $N_1+N_2$ even and $\kappa=1/2$ for $N_1+N_2$ odd.
This effectively allows us to exchange the order of the sums in the
above equation. Then by comparing the resulting equation
with~(\ref{tr1}), we obtain

\begin{align}
\nu(N_1+N_2,J)=\sum_{j_1,m_1}\sum_{j2,m2}\nu(N_1,j_1)\nu(N_2,j_2)(-1)^{2(j_1-j_2+J)}(2J+1)\nonumber\\
\times
\Bigl\{\begin{pmatrix}j_1&&j_2&&J\\m_1&&m_2&&-J\end{pmatrix}\Bigl\}^2.\label{main}
\end{align}
Herein,   we have replaced $M$ by its maximum value $J$ (or
equivalently by $-J$ because of the symmetry) since the sum does not
depend on this quantum number; once again the condition $m_1+m_2=J$
is implied.

Equation~(\ref{main}) can be regarded as a decomposition law for the
degeneracy; many useful relations satisfied by  the latter  can be
easily obtained from it.  Let us first begin by noting that
\begin{align}
\sum_{J=\kappa}^{\frac{N}{2}}\nu(N,J)&=\binom{N}{\frac{N}{2}-\kappa},\label{pr1}\\
\sum_{J=\kappa}^{\frac{N}{2}}(2J+1)&\nu(N,J)=2^N.\label{pr2}
\end{align}
The first equation can be readily proved by expanding the sum over
$J$. The second one simply expresses the fact that the sum of the
dimensions of the subspaces $\mathbb C^{2j+1}$ is equal to the
dimension of the total state space, $\mathbb C^{2\otimes N}$.
Furthermore, if we let $J$ to take the value $\frac{N_1+N_2}{2}$ in
equation (\ref{main}), we  obtain
\begin{align}
(-1)^{N_1+N_2}(N_1+N_2+1)\sum_{j_1,m_1}\sum_{j2,m2}\nu(N_1,j_1)\nu(N_2,j_2)(-1)^{2(j_1-j_2)}\nonumber\\
\times
\Bigl\{\begin{pmatrix}j_1&&j_2&&\frac{N_1+N_2}{2}\\m_1&&m_2&&-\frac{N_1+N_2}{2}\end{pmatrix}\Bigl\}^2=1.
\end{align}

Now let us suppose that $J=0$, which is possible only when  $N_1$
and $N_2$ are either both even or both odd positive integers. Here
it should be noted that the denominator of  the corresponding Wigner
$3j$-symbol contains the product $(j_1-j_2)!(j_2-j_1)!$~\cite{14};
but since $x!=\infty$ for $x<0$, we conclude that when $J=0$, the
quantity under the sum sign in the right-hand side of equation
(\ref{main}) is nonzero only when $j_1=j_2$. In fact one should
have~\cite{12,14}
\begin{equation}
\begin{pmatrix}j_1&&j_2&&0\\m_1&&m_2&&0\end{pmatrix}=(-1)^{j_1-m_1}\sqrt{\frac{1}{2j_1+1}}\delta_{j_1j_2}\delta_{-m_1m_2}.\end{equation}
By inserting the latter expression of  Wigner $3j$-symbol into
equation~(\ref{main}), and performing the sum over $j_2$ and $m_2$,
we obtain
\begin{align}
\nu(N_1+N_2,0)&=\sum_{j}^{\min\{\frac{N_1}{2},\frac{N_2}{2}\}}\sum_{m=-j}^{j}\nu(N_1,j)
\nu(N_2,j)\frac{(-1)^{2(j-m)}}{2j+1}\nonumber \\
&=\sum_{j}^{\min\{\frac{N_1}{2},\frac{N_2}{2}\}}\nu(N_1,j)
\nu(N_2,j),
\end{align}
where we have used the fact that
$\sum_{m=-j}^{j}(-1)^{2m}=(-1)^{2j}(2j+1)$. It immediately follows
that
\begin{equation}
\sum_{j}^{N/2}\nu(N,j)^2=\frac{(2N)!}{(N+1)(N!)^2}.\end{equation}
The above procedure can be easily generalized to further
decompositions of the total number of spins.

  \subparagraph*{Dynamics of
two qubits in separate spin baths.} As a second application, let us
investigate the dynamics of two qubits coupled via ising
interactions to separate spin environments of the same size, $N$.
The total angular momentum operators of the latter are denoted by
$\vec{J}$ and $\vec{\mathcal J}$. The full Hamiltonian of the
composite system is given by
\begin{equation}
H=\lambda(\sigma^1_x\sigma^2_x+\sigma^1_y\sigma^2_y)+\delta\sigma^1_z\sigma^2_z+\frac{\gamma}{\sqrt{N}}(\sigma^1_z
J_z+\sigma^2_z\mathcal
J_z)+\mu(\sigma^1_z+\sigma^2_z)+H_{B_1}+H_{B_2}.\end{equation} Here,
$\lambda$ and $\delta$ are the strengths of interaction of the
central qubits with each other, $\gamma$ is the coupling constant to
the baths, and $\mu$ is the strength of an applied magnetic field.
The operators $H_{B_i}$, with $i=1,2$, denote the  Hamiltonians of
the spin baths. One can show that
 the interaction Hamiltonian
describing the coupling of the central qubits to the environments is
diagonal in the standard basis of $\mathbb C^2\otimes \mathbb C^2$,
namely,
\begin{equation}
H_I=\frac{\gamma}{\sqrt{N}}
diag(-\Sigma_z,-\Delta_z,\Delta_z,\Sigma_z),\end{equation} where we
have introduced the operators
$\vec{\Sigma}=\vec{J}+\vec{\mathcal{J}}$ and
$\vec{\Delta}=\vec{J}-\vec{\mathcal{J}}$. Then it can be shown that
the model Hamiltonian is given by the direct sum of the Hamiltonian
operators $H_1$ and $H_2$, where
\begin{subequations}
\begin{equation}
H_1=\sigma_z(2\mu+\frac{\gamma}{\sqrt{N}}\Sigma_z)+\mathbb I_2
(H_B+\delta),
\end{equation}
\begin{equation}
H_2=2\lambda\sigma_x+\frac{\gamma}{\sqrt{N}}\sigma_z\Delta_z+\mathbb
I_2(H_B-\delta),
\end{equation}
\end{subequations}
with  $H_B=H_{B_1}+H_{B_2}$ and $\mathbb I_2$ is the $2 \times 2$
unit matrix.  Note that the basis vectors of the subspace
corresponding to $H_1$ are given by
\begin{eqnarray}
|\downarrow\rangle&\equiv|--\rangle,\\
|\uparrow\rangle&\equiv|++\rangle;
\end{eqnarray}
 those associated with $H_2$ are given by
\begin{eqnarray}
|0\rangle&\equiv|-+\rangle,\\
|1\rangle&\equiv|+-\rangle.
\end{eqnarray}
 Thus the system under consideration can be mapped
onto two pseudo two-level systems $\mathbf S_1$ and $\mathbf S_2$
whose dynamics is governed by the operators $H_1$ and $H_2$,
respectively. Each one  is coupled to a spin environment consisted
of $2N$ spin-$\frac{1}{2}$ particles with the only exception that
$\mathbf{S}_1$ and $\mathbf S_2$ see different compositions of the
total angular momentum, namely  $\Sigma_z$ and $\Delta_z$,
respectively. Notice that the above pseudo systems become completely
independent from each other if the  initial density matrix of the
qubits takes the form
\begin{equation}
\rho(0)=\begin{pmatrix}\rho_{11}^0&&0&&0&&\rho_{14}^0\\0&&\rho_{22}^0
&&\rho_{23}^0&&0\\0&&\rho_{32}^0
&&\rho_{33}^0&&0\\\rho_{41}^0&&0&&0&&\rho_{44}^0\end{pmatrix}.\end{equation}
 In such a case, it is sufficient to investigate the coupling of
each pseudo system separately. For a reason that will become
apparent bellow, we set
 $H_B=H_{B_1}+H_{B_2}=h(J_z-\mathcal{J}_z)$, where $h$ is the
strength of an  applied magnetic field. Moreover, we assume that the
baths are initially in thermal equilibrium at temperatures
$T_1=T_2=T$ (we set $k_B=1$); the corresponding total initial
density matrix is  given by
\begin{equation}
\rho_{B}(0)=\exp(-h\beta\Delta_z)/\Bigl[2
\cosh\Bigl(\frac{h\beta}{2}\Bigl)\Bigl]^{2N},\end{equation} where
$\beta=1/T$ is the inverse temperature and $Z=\Bigl[2
\cosh\Bigl(\frac{h\beta}{2}\Bigl)\Bigl]^{2N}$ is the partition
function. Under the above assumptions, the contributions of the
coupling constant $\delta$ can be neglected.

The dynamics of  $\mathbf S_2$ is quite trivial since the
corresponding  time evolution operator is diagonal. Indeed, it is
easy to show that $\rho_{11}(t)=\rho_{11}^0$ and
$\rho_{44}(t)=\rho_{44}^0$. Moreover,
\begin{align}\rho_{14}(t)&=Z^{-1}\rho_{14}^0\sum_{j_1,m_1}\sum_{j_2,m_2}\nu(N,j_1)\nu(N,j_2)
\nonumber\\&\times
\exp\{2i[2\mu+\gamma(m_1+m_2)/\sqrt{N}]t-h\beta(m_1-m_2)\}.\label{expg}\end{align}
In the special case when $h=0$ or $T\to\infty$, we can write
\begin{align}
\rho_{14}(t)&=2^{-2N}\rho_{14}^0e^{4it\mu}\sum_{J,M}\nu(2N,J)
e^{2\sqrt{2}it\gamma M/\sqrt{2N}}\nonumber \\
&=\rho_{14}^0e^{4it\mu}\cos\bigl(\frac{\gamma
t}{\sqrt{N}}\Bigl)^{2N}.
\end{align}
For arbitrary values of $h$ and $T$, the right-hand side of
equation~(\ref{expg}) can be evaluated within the computational
basis; this  yields
 \begin{equation}
\rho_{14}(t)/\rho_{14}^0=e^{4it\mu}\Bigl[1+\frac{\cos^2(\gamma
t/\sqrt{N})-1}{\cosh^2(h\beta/2)}\Bigl]^N.\end{equation} Then, by
expanding the cosine function in Taylor series and taking the limit
$N\to\infty$, we obtain the Gaussian decay law:
\begin{equation}
 |\frac{\rho_{14}(t)}{\rho_{14}^0}|=
\exp\Bigl\{-\frac{\gamma^2t^2}{\cosh^2(h\beta/2)}\Bigl\}.\label{gdec}\end{equation}
This means that the decoherence time scale is  given by
$\tau_D=\cosh(h\beta/2)/|\gamma|$. Obviously $\tau_D\to\infty$ as $T
\to 0$ or $h \to \infty$.

\begin{figure}[htba]
 {\centering
{\resizebox*{0.7\textwidth}{!}{\includegraphics{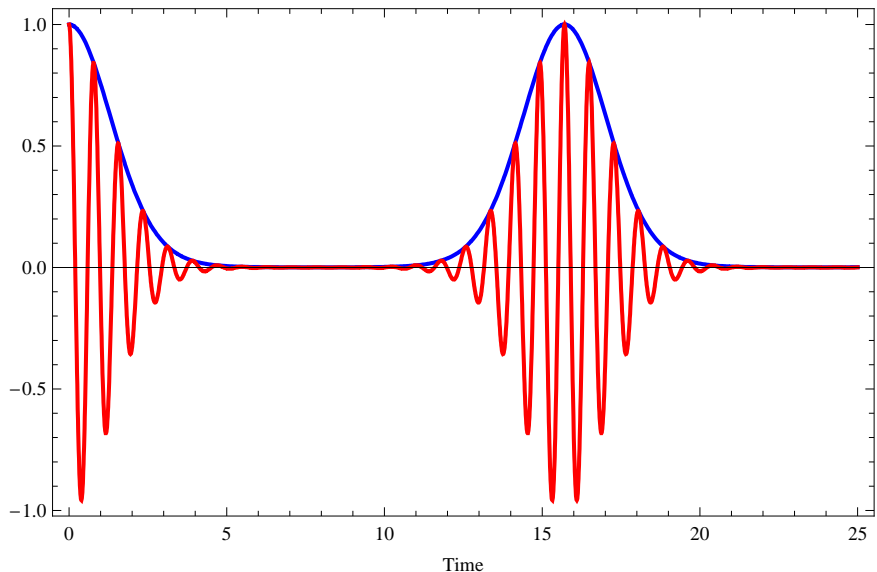}}
\par}
\caption{ (Color online) \label{figure1}Evolution in time of the
real part of $\rho_{14}(t)/\rho^0_{14}$ (oscillating curve) and the
concurrence (enveloping curve) corresponding to the initial state
$(|--\rangle+|++\rangle)/\sqrt{2}$. Here, $N=100$, $\gamma=2$,
 $h\beta=1$, and $\mu = 4$. For $t<10$, the curves coincide with those of the limit $N\to\infty$.}}
\end{figure}

  As a measure
of entanglement, we use the concurrence defined by~\cite{15}
\begin{equation}
C(\rho)=\max\{0,2\max[\sqrt{\lambda_i}]-\sum_{i=1}^4\sqrt{\lambda_i}\},\end{equation}
where the quantities  $\lambda_i$ are the eigenvalues of the
operator
$\rho(\sigma_y\otimes\sigma_y)\rho^*(\sigma_y\otimes\sigma_y)$. In
our case, when applied to $\rho(t)$, the above definition of the
concurrence leads to the evaluation of the eigenvalues of the
operator $\rho(t)\sigma_x\rho(t)^*\sigma_x$ where  $\rho(t)$ is now
restricted to the subspace of $H_1$. A straight forward calculation
yields
\begin{equation}
C(t)=2|\rho_{14}(t)|.\end{equation} An example of the evolution in
time of the real value  of  $\rho_{14}(t)$ along with the
concurrence $C(t)$ corresponding to the initial state
$(|--\rangle+|++\rangle)/\sqrt{2}$ is shown in figure~\ref{figure1}.
We notice the revival of the concurrence in the case of finite
number of spins. At short times, the curves corresponding to
$N\to\infty$ coincide with those of finite $N$.

It should be stressed that when the Hamiltonian of the composite
spin bath is given by $H_B=h(J_z+\mathcal{J}_z)=h\Sigma_z$, then
\begin{equation}
\rho_{14}(t)=\rho_{14}^0 e^{4i\mu t}\Bigl[\cos(\gamma
t/\sqrt{N})-i\sin(\gamma
t/\sqrt{N})\tanh(h\beta/2)\Bigl]^{2N}.\end{equation} The existence
of the sine function makes it not possible to find a relation
similar to (\ref{gdec}) when $N\to\infty$. However if we rescale the
coupling constant $\gamma$ by $N$ instead of $\sqrt{N}$, that
is~\cite{16},
\begin{equation}
\frac{\gamma}{\sqrt{N}}\rightarrow\frac{\gamma}{N},\end{equation}
exact analytical expression can be derived for the case of an
infinite number of spins, namely,
\begin{equation}
\rho_{14}(t)=\rho_{14}^0 \exp\Bigl\{-it[4\mu+\gamma
\tanh(h\beta/2)]\Bigl\}.\label{coh}\end{equation} Consequently the
central qubits preserve their coherence, since the decoherence time
scale in this case is infinite, as indicated by formula~(\ref{coh}).
With the new scaling of $\gamma$,  the larger the number of spins to
which the qubits are coupled, the less appreciable is the
decoherence.

 The
Hamiltonian operator $H_2$ can be diagonalized by dealing with the
operator $\Delta_z$ as a scalar. This yields the following matrix
elements in $\mathbb C^2$:
\begin{align}
U_{22}(t)&=\cos\Bigl(t\sqrt{4\lambda^2+\gamma^2\Delta_z^2/N}\Bigl)+i\frac{\gamma}{\sqrt{N}}\Delta_z\frac{\sin
\Bigl(t\sqrt{4\lambda^2+\gamma^2\Delta_z^2/N}\Bigl)}{\sqrt{4\lambda^2+\gamma^2\Delta_z^2/N}}\\
U_{23}(t)&=U_{32}(t)=-\frac{2i\lambda} {\sqrt{4\lambda^2+\gamma^2\Delta_z^2/N}}\sin\Bigl(t\sqrt{4\lambda^2+\gamma^2\Delta_z^2/N}\Bigl)\\
U_{33}(t)&=\cos\Bigl(t\sqrt{4\lambda^2+\gamma^2\Delta_z^2/N}\Bigl)-i\frac{\gamma}{\sqrt{N}}\Delta_z\frac{\sin
\Bigl(t\sqrt{4\lambda^2+\gamma^2\Delta_z^2/N}\Bigl)}{\sqrt{4\lambda^2+\gamma^2\Delta_z^2/N}},
\end{align}
Here we have omitted the contribution of $H_B=h \Delta_z$ since it
simply introduces a global unitary term to the dynamics.

\begin{figure}[htba]
 {\centering
{\resizebox*{0.7\textwidth}{!}{\includegraphics{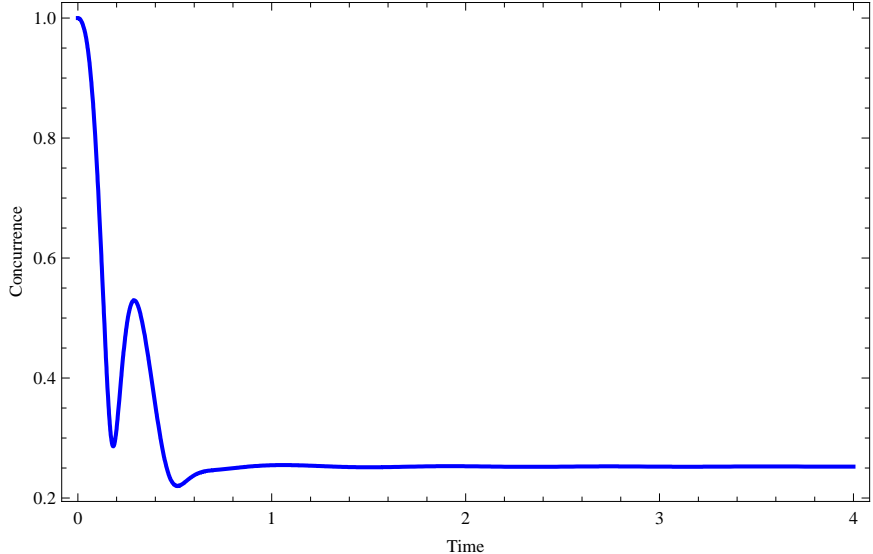}}
\par}
\caption{ (Color online) \label{figure2} Concurrence as a function
of time in  the case of the initial state
$(|-+\rangle+|+-\rangle)/\sqrt{2}$ for  $N=100$, $\gamma=4$,
 $h\beta=4$, and $\lambda = 2$.}}
\end{figure}

\begin{figure}[htba]
 {\centering
{\resizebox*{0.7\textwidth}{!}{\includegraphics{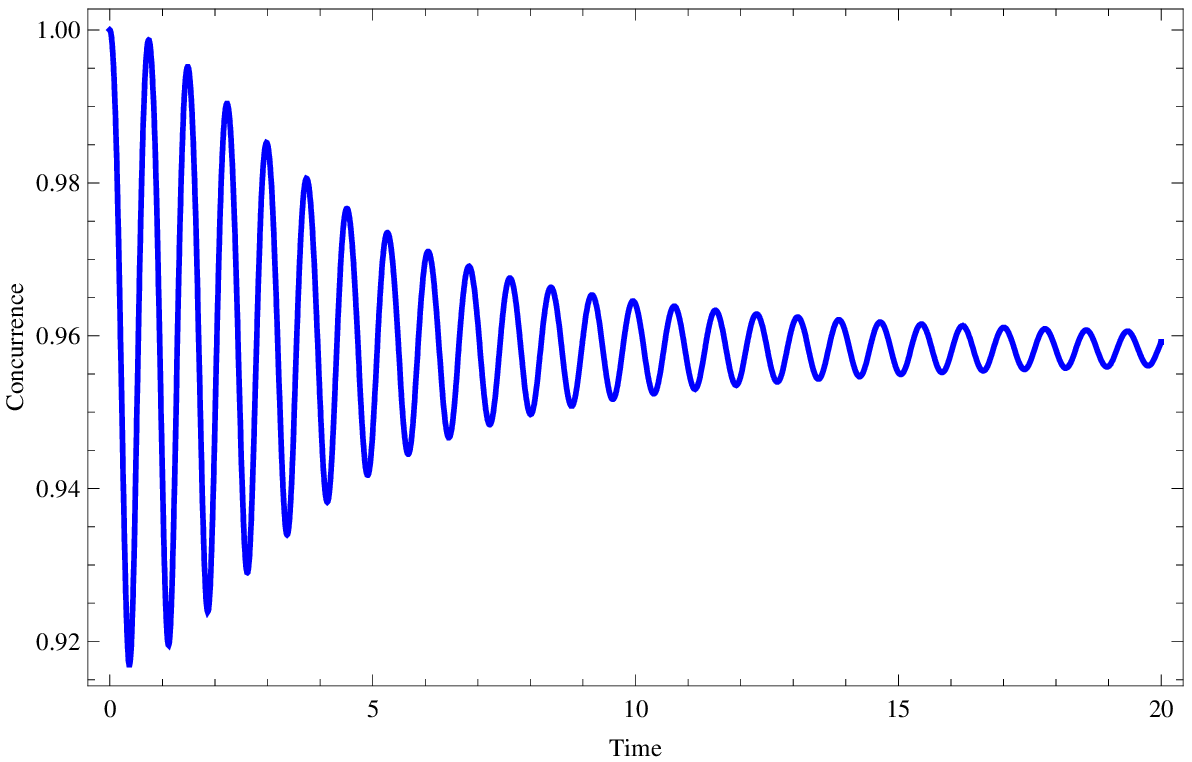}}
\par}
\caption{ (Color online) \label{figure3} Concurrence as a function
of time in  the case of the initial state
$(|-+\rangle+|+-\rangle)/\sqrt{2}$ for  $N=100$, $\gamma=1$,
 $h\beta=1$, and $\lambda = 2$.}}
\end{figure}

Let us consider the case when the qubits are initially prepared in
the  maximally entangled state
$|\psi\rangle=\frac{1}{\sqrt{2}}(|-+\rangle+|+-\rangle)$.( the case
of the singlet state displays a similar behavior.) Clearly, the
density matrix  $\rho(0)=|\psi\rangle\langle\psi|$ belongs to the
subspace corresponding to the Hamiltonian $H_2$. Using the fact that
$|U_{22}(t)|^2+|U_{23}(t)|^2=\mathbb{I}_{B}$, and $U_{22}(t)
U_{23}^\dag(t)+U_{23}(t)U_{33}^\dag=0$, it can be shown that the
elements of the above density
 matrix evolve in time according to
$\rho_{22}(t)=\frac{1}{2}[1-g(t)]$, $\rho_{23}=\frac{1}{2}[1-f(t)]$,
where
\begin{align}
g(t)=\frac{4\lambda\gamma}{[2
\cosh(h\beta/2)]^{2N}}\mathrm{tr}\Bigl\{ \frac{\Delta_z
e^{-h\beta\Delta_z}}{\sqrt{N}}
\frac{\sin^2\Bigl(t\sqrt{4\lambda^2+\gamma^2\Delta_z^2/N}\Bigl)}
{4\lambda^2+\gamma^2\Delta_z^2/N}\Big\},
\end{align}
and
\begin{align}
f(t)=\frac{1}{[2 \cosh(h
\beta/2)]^{2N}}&\mathrm{tr}\Bigl\{\frac{2\gamma^2 \Delta_z^2
e^{-h\beta\Delta_z}}{N}\frac{\sin^2\Bigl(t\sqrt{4\lambda^2+\gamma^2\Delta_z^2/N}\Bigl)}
{4\lambda^2+\gamma^2\Delta_z^2/N}\nonumber\\&-i\frac{\gamma
e^{-h\beta\Delta_z}} {\sqrt{N}}\Delta_z \frac{\sin\Bigl(2
t\sqrt{4\lambda^2+\gamma^2\Delta_z^2/N}\Bigl)}
{\sqrt{4\lambda^2+\gamma^2\Delta_z^2/N}}\Bigl\}.
\end{align}
Figures~\ref{figure2} and \ref{figure3} display the behavior of the
concurrence as a function of time for some particular values of the
model parameters. We can see that for $h\beta=1$ ( i.e. at
relatively high temperature) the concurrence  shows damped
oscillations and converges to a certain asymptotic limit which can
be analytically derived, as we shall see bellow, only for $h=0$
and/or $\beta=0$. As $h\beta$ increases, the oscillations disappear
and the concurrence converges to lower asymptotic values as shown in
figure~\ref{figure2}.

In what follows we focus our attention on the infinite temperature
limit, i.e, $\beta\to 0$. In this case the reduced  density matrix
takes the form
\begin{align}
\rho(t)=\frac{1}{2}\begin{pmatrix}1&&1-f(t)\\1-f(t)&&1\end{pmatrix},\label{time}\end{align}
whereas  the function $f(t)$ simplifies to \begin{equation}
f(t)=2^{-2N}\mathrm{tr}\Bigl\{\frac{2\gamma^2
\Delta_z^2}{N}\frac{\sin^2\Bigl(t\sqrt{4\lambda^2+\gamma^2\Delta_z^2/N}\Bigl)}
{4\lambda^2+\gamma^2\Delta_z^2/N}\Big\}.\end{equation} Notice that
$0 \le f(t) \le 2$ , in accordance with the general properties of
density matrices in $\mathbb C^2$. This enables us to derive the
following explicit expression for the concurrence:
\begin{align}
C(t)=  &\frac{1}{2}[\sqrt{f(t)^2-4f(t)+4}-f(t)]\nonumber\\
=&1-f(t)\label{con}.
 \end{align}
In the thermodynamic limit, $N\to\infty$, the function $f(t)$ can be
expressed as
\begin{equation}
f(t)=4\gamma^2 \sqrt{\frac{2}{\pi}}\int_{-\infty}^{\infty}\frac{x^2
e^{-2 x^2}}{4\lambda^2+2\gamma^2 x^2}
\sin^2\Bigl(t\sqrt{4\lambda^2+2\gamma^2 x^2}\Bigl)
dx.\label{inf}\end{equation} Some comments are in order here: We
have shown in~\cite{9} that the operator $J_z/\sqrt{N}$ converges to
a real normal random variable $\alpha$ with the probability density
function $F(\alpha)=\sqrt{2/\pi}\exp\{-2\alpha^2\}$; this is also
the case for the operator $\mathcal{J}_z/\sqrt{N}$. Thus we are led
to the task of finding the probability distribution function
$L(\alpha)$ of the sum of two independent random variables
$\alpha_1$ and $\alpha_2$ characterized by $F(\alpha_1)$ and
$F(\alpha_2)$, respectively. (note that the probability distribution
function of $a\alpha$, where $a$ is nonzero real number, is equal to
$ (1/|a|) F(\alpha/a)$.) The function $L(\alpha)$ is simply given by
the convolution of $F(\alpha)$ with itself, which yields
$L(\alpha)=(1/\sqrt{\pi})\exp\{-\alpha^2\}$. This becomes apparent
from the change of variable $\alpha\to\sqrt{2}\alpha$ carried out in
equation~(\ref{inf}). An other way to see that is to simply notice
that $\Delta_z/(\sqrt{2N})$ converges to the random variable
$\alpha\mapsto F(\alpha)$. From equation~(\ref{inf}) it follows that
\begin{equation}
\lim_{t\to\infty} f(t)=1-2\sqrt{\pi}\frac{\lambda}{\gamma}
e^{4\frac{\lambda^2}{\gamma^2}}
\mathrm{erfc}\Bigl(2\frac{\lambda}{\gamma}\Bigl),\end{equation}
where $\mathrm{erfc} (x)$ denotes the complementary error function.
By virtue of equation~(\ref{con}), we obtain
\begin{equation}
C(\infty)=\lim_{t\to\infty}C(t)= 2\sqrt{\pi}\frac{\lambda}{\gamma}
e^{4\frac{\lambda^2}{\gamma^2}}
\mathrm{erfc}\Bigl(2\frac{\lambda}{\gamma}\Bigl).\label{coninf}\end{equation}
\begin{figure}[htba]
 {\centering
{\resizebox*{0.7\textwidth}{!}{\includegraphics{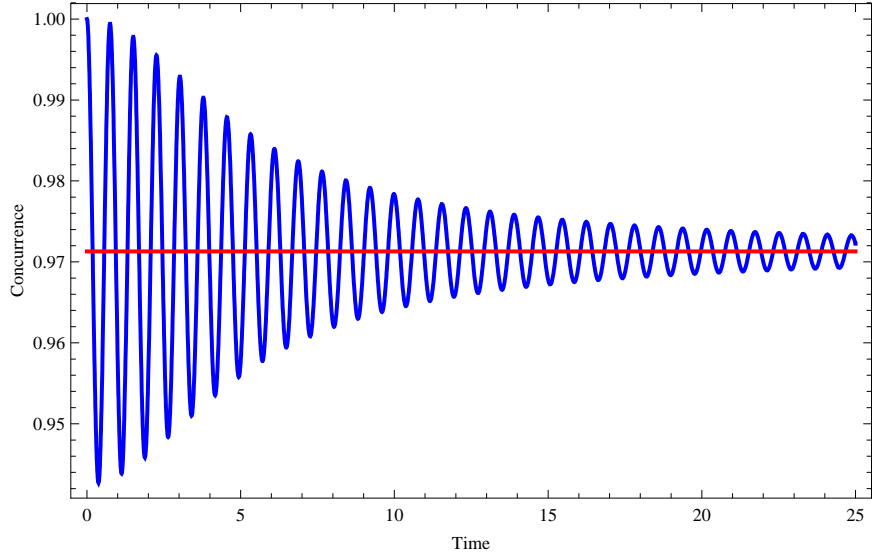}}
\par}
\caption{ (Color online) \label{figure4} Concurrence as a function
of time in  the case of the initial state
$(|-+\rangle+|+-\rangle)/\sqrt{2}$ for  $N=100$ (coincides with that
of the limit $N\to\infty$), $\gamma=1$,
 $h\beta=0$, and $\lambda = 2$. The straight line corresponds to the asymptotic limit.}}
\end{figure}
\begin{figure}[tba]
 {\centering
{\resizebox*{0.7\textwidth}{!}{\includegraphics{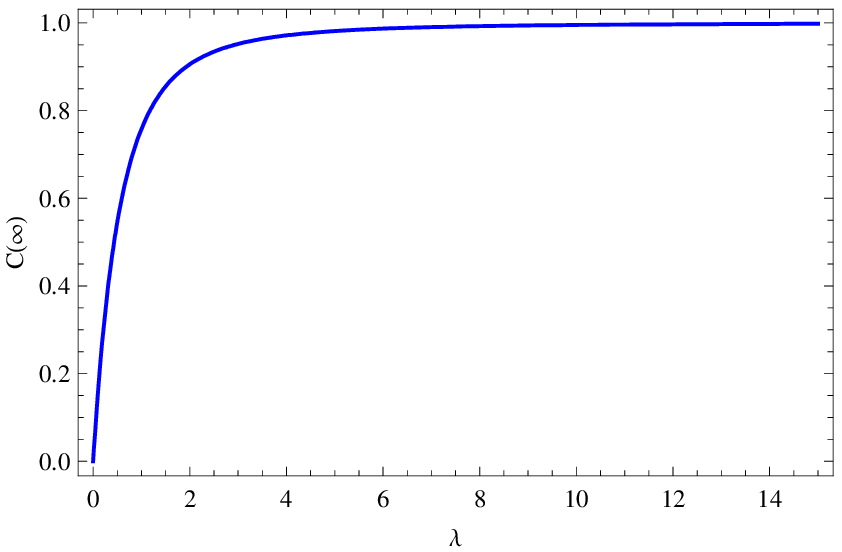}}
\par}
\caption{ (Color online) \label{figure5} $C(\infty)$ as a function
of $\lambda$ for $\gamma=2$.}}
\end{figure}

\begin{figure}[ta]
 {\centering
{\resizebox*{0.7\textwidth}{!}{\includegraphics{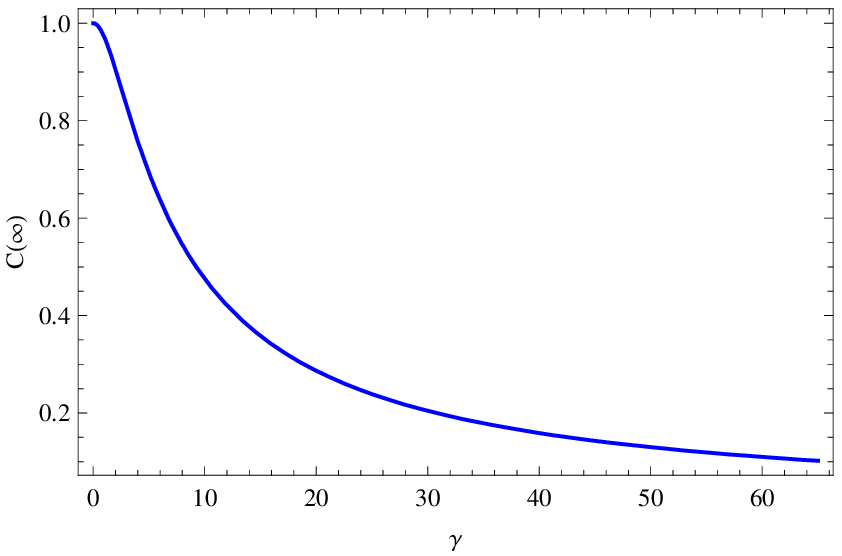}}
\par}
\caption{ (Color online) \label{figure6} $C(\infty)$ as a function
of $\gamma$ for  $\lambda=2$.}}
\end{figure}
 In figure~\ref{figure4} we have plotted the concurrence as a function
of time in the limit $N\to\infty$ along with the asymptotic value
given by formula~(\ref{coninf}). The behavior of $C(\infty)$ as a
function of $\lambda$ and $\gamma$ is shown in figures~\ref{figure5}
and ~\ref{figure6} . As one may expect,
$\lim\limits_{\lambda\to\infty}C(\infty)=1$, and
$\lim\limits_{\gamma\to\infty}C(\infty)=0$. This confirms the
results of~\cite{10} where it is shown that strong coupling between
the central qubits reduces the effect of the environment on their
dynamics. Finally it is worth mentioning that due to the $XY$
interaction between the central spins, entanglement will be
generated between them when the initial state is $|\pm\mp\rangle$.
However, the corresponding off-diagonal elements of the reduced
density matrix vanish at long times, making the asymptotic state of
the qubits unentangled.

In summary we have used the invariance of the trace to derive
analytical properties of the degeneracy $\nu(N,j)$, and to describe
the dynamics of two qubits embedded in separate spin baths. We have
shown that when the baths have the same size, the form of the model
Hamiltonian enables us to map the full  dynamics onto the evolution
in time of two pseudo two-level systems coupled to a spin bath whose
size is twice larger than the physical ones. This allowed us to
derive the limit of an infinite number of spins within the
environments and to analytically calculate the asymptotic state. The
results of this work provide more evidences regarding the role
played by the mutual interactions between the central qubits in
diminishing the effects of their coupling to the surrounding spin
environments.


\begin{thebibliography}{99}
\bibitem{1}
H.~P.~Breuer and F.~Petruccione, The Theory of Open  Quantum
Systems, Oxford University Press, Oxford, 2002.
\bibitem{2}
W.~H.~Zurek, Phys.\ Today 44 (1991) No. 10, 36
\bibitem{3}
D.~P.~DiVincenzo, D.~Loss, J. Magn. Magn. Matter. 200 (1999) 202 .
\bibitem{4}
W.~H.~Zurek, Rev.\ Mod.\ Phys. 75 (2003) 715-775 .
\bibitem{5}
W.~Zhang, N.~Konstantinidis, K.~Al-Hassanieh,  V.~V.~Dobrovitski,
J.\ Phys.:\ Condens.\ Matter 19  (2007) 083202 .
\bibitem{6}
M.~A.~Nielsen, I.~L.~Chuang, Quantum Computation and Quantum
Information, Cambridge University Press, Cambridge, 2000.
\bibitem{7}
D.~ Loss,
 D.~P.~ DiVincenzo Phys.\ Rev.\ A 57 (1998) 120.
\bibitem{8}
G.~Burkard, D.~Loss,  D.~P.~DiVincenzo , Phys.\ Rev.\ B 59 (1999)
2070.
\bibitem{9}
Y.~Hamdouni,  F.~Petruccione, Phys.\ Rev.\ B  76 (2007) 174306
\bibitem{10}
 Y.~Hamdouni, M.~Fannes,
F.~Petruccione, Pyhs.\ Rev.\ B 73 (2006) 245323; X.~Z.~Yuan,
H.~S.~Goan,  K.~D.~Zhu, Phys.\ Rev.\ B 75 (2007) 045331;
 Y.~Hamdouni, J.\ Phys.\ A:\ Math.\ Theor. 40(2007) 11569 ; Y.~Hamdouni, J.\ Phys.\ A:\ Math.\
Theor. 41 (2008) 135302.
\bibitem{11}
X.~Wang,~ K.~ M{\o}lmer, Eur.\ Phys.\ J.\ D  (2002) 385.
\bibitem{12}
C. Cohen-Tannoudji, B. Diu, and F. Laloë,  Quantum Mechanics, Wiley,
New York, 1977, Vols. I and II.
\bibitem{13}
W.~ Von Waldenfels, S\'eminaire de probabilit\'e (Starsburg),
tome(24), pp.349-356, Springer-Verlag, Berlin, 1990.
\bibitem{14}
L.~D.~Landau, and E.~M.~Lifchitz, Quantum Mechanics, Pergamon Press,
Oxford, 1976.
\bibitem{15}
W.~K.~Wootters Phys.\ Rev.\ Lett.~80  (1998) 2245.
\bibitem{16}
H.~ T.~Quant , Z.~D.~Wang,  C.~P.~Sun   Phys. Rev. A  76  ( 2007)
012104.
\end{thebibliography}
\end{document}